\documentclass[aps,pra,twocolumn,longbibliography,footinbib,groupedaddress]{revtex4-1}
\usepackage{graphicx}
\usepackage{multirow}
\usepackage{amsmath}
\usepackage{amssymb}
\usepackage[utf8]{inputenc}
\usepackage{array}
\usepackage{color}
\usepackage[caption=false]{subfig}

\usepackage{ulem}

\begin{document}

\title{Diffusing-Wave Spectroscopy of Cold Atoms in Ballistic Motion}

\author{Aur\'elien Eloy$^{1}$}
\author{Zhibin Yao$^{1}$}
\author{Romain Bachelard$^{2}$}
\author{William Guerin$^1$}
\author{Mathilde Fouch\'e$^1$}  \email{mathilde.fouche@inphyni.cnrs.fr}
\author{Robin Kaiser$^1$}

\affiliation{$^{1}$Universit\'e C\^ote d'Azur, CNRS, INPHYNI, France}
\affiliation{$^{2}$Departamento de F\'{\i}sica, Universidade Federal de S\~{a}o Carlos, Rod. Washington Lu\'{\i}s, km 235 - SP-310, 13565-905 S\~{a}o Carlos, SP, Brazil}

\date{\today}

\begin{abstract}
{Diffusing-wave spectroscopy is a powerful technique which consists in measuring the temporal correlation function of the intensity of light multiply scattered by a medium. In this paper, we apply this technique to cold atoms under purely ballistic motion and we investigate the transition between the single and the multiple-scattering regime. The intensity correlation function changes from a simple Gaussian function, whose width reveals the sample temperature, to a more complex decay shortened by the frequency redistribution in the thick cloud. These features are quantitatively compared to simulations with a phase-coherent and an incoherent model. Both show a very good agreement with the experiments.}

\end{abstract}


\maketitle


\section{Introduction} \label{Sec:Intro}

The study of fluctuations and correlations often gives access to information not contained in averaged values. Among the many statistical properties of a fluctuating field, the intensity correlation function is largely used in a number of areas, from astronomy~\cite{HBT:1956,HBT:nous}, to quantum optics~\cite{Glauber:1963,Arecchi:1966,Book:Loudon,Bachor:book}, particle physics~\cite{Alexander:2003}, and to mesoscopic optics. In the latter, it has been applied to the fluctuations of light scattered by a disordered medium. First used in the single-scattering regime with a technique known as dynamic light scattering or quasielastic light scattering\,\cite{Clark1970, Berne1976}, it was then extended to strong multiple-scattering regime. This powerful technique, called diffusing-wave spectroscopy (DWS)\,\cite{Maret1987, Pine1988, Pine1990}, allows probing extremely small displacement, much smaller than the optical wavelength. It has found a wide range of applications, going from particle sizing\,\cite{Pine1990,Rega2001, Scheffold2002}, Brownian motion\,\cite{Weitz1989, Kao1993}, hydrodynamic interactions\,\cite{Fraden1990, Weitz1993}, microscopic dynamics\,\cite{Hebraud1997, Menon1997} and medical applications\,\cite{Durduran2004, Ninck2001}.

DWS consists in measuring the temporal intensity correlation $g^{(2)}(\tau)$ of the light scattered by the medium under investigation.
In the case of a stationary process, this quantity is expressed as:
\begin{equation}
g^{(2)}(\tau) = \frac{\langle I(t)I(t + \tau) \rangle}{\langle I(t) \rangle^2},
\end{equation}
\noindent
where $\langle\rangle$ denotes time averaging. If light is scattered by a large number of particles it can be considered as chaotic light for which $g^{(2)}(\tau = 0)$ is equal to 2, representing photon bunching, and it goes to 1 for large $\tau$ when events become uncorrelated. The temporal decay of $g^{(2)}(\tau)$ gives access to the mean square displacement of the scatterers. On the other hand, the shape of $g^{(2)}(\tau)$ on different time scales provides information on the scatterers dynamics\,\cite{Stephen1988, Weitz1989, Menon1997}. The standard theoretical approach to describe DWS is to use the diffusion theory for light. The first order correlation function $g^{(1)}(\tau)$ can be derived\,\cite{Pine1990}, and the second order is obtained using the Siegert relation\,\cite{Book:Loudon, Pine1990}:
\begin{equation}
g^{(2)}(\tau) = 1 + \beta|g^{(1)}(\tau)|^2.
\label{eq:Siegert}
\end{equation}
The factor $\beta$ is related to the number of detected spatial and polarization modes. For a detector
whose radius is much smaller than the spatial coherence length of the scattered light and for polarized light
$\beta$ is equal to unity\,\cite{Book:Loudon}.

Here, we apply the DWS technique on a cold atomic cloud. The light beam is set close to resonance and its frequency is changed to probe the transition from the single to the multiple-scattering regime. This is a first difference from other DWS measurements where light is always considered to be far from any resonance. Cold atoms also allow for a pure ballistic regime, which is usually hardly reached in standard DWS measurements due to the surrounding fluid in which scatterers are immersed\,\cite{Maret1987, Pine1988, Weitz1989, Kao1993}. The intensity correlation function of light scattered by cold atoms has already been measured in several experiments and signatures of the Doppler broadening, Lamb-Dicke narrowing, radiation trapping and inelastic scattering (Mollow triplet) could be obtained~\cite{Article:Bali1996,Jurczak:1996,Stites2004,Nakayama2010,Grover2015,MuhammedShafi2016}. However, these measurements were done with fluorescence from atoms trapped in a magneto-optical trap (MOT) or during a molasse phase, with thus large beams (probing different parts of the cloud and thus different optical thicknesses) and quite strong intensities, leading to a complicated interplay between all these effects.

In this paper, we perform intensity correlation measurements while the atoms are ballistically expanding with a small and weak probe beam, which we can tune independently from the laser-cooling beams. At the cost of a long integration time, this allows us to avoid inelastic effects and precisely investigate the transition from the single-scattering to the multiple-scattering regime with a well-defined optical thickness. The behavior of the $g^{(2)}$ function is distinctively different from that obtained in other media (scatterers in a fluid) and it is only related to the velocity distribution of the atoms. We observe in particular a change of shape of the $g^{(2)}$ function, from a Gaussian in the single-scattering regime to a more complex shape in the multiple-scattering regime, which in good agreement with simulations.


\section{Experimental setup} \label{sec:SetUp}

\begin{figure}\centering
	\includegraphics[width = 86mm]{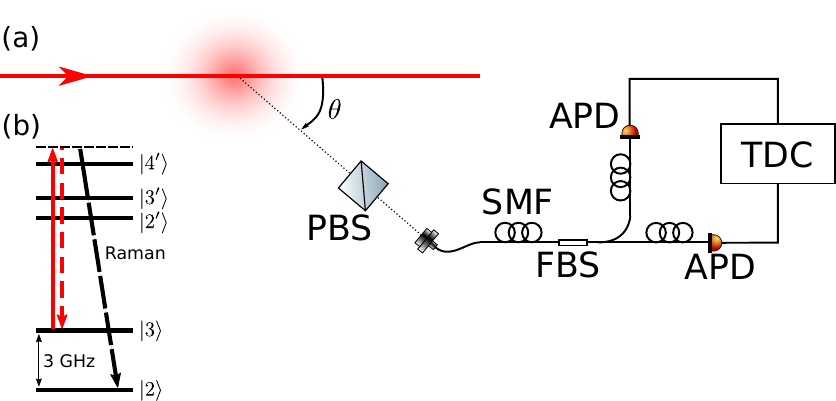}
	\caption{Experimental set-up. Light scattered in the MOT is collected using a single-mode fiber (SMF), after
	passing through a polarizing beam splitter (PBS) to select only one polarization.
	The light is then split in a 50:50 fiber beam splitter (FBS). The outputs of the FBS are connected to two
	APDs. The single counts in each APDs are time-tagged by a time-to-digital converter (TDC) and analyzed with a
	computer. Left bottom: Relevant atomic levels of the D$_2$ hyperfine transition of $^{85}$Rb.}
      \label{fig:setup}
\end{figure}

In our experiment, the scattering sample is a cold atomic vapor produced by a MOT that can contain up to a few $10^{9}$ atoms of $^{85}\mathrm{Rb}$ with a Gaussian density distribution of rms size $R \simeq 1 \, \mathrm{mm}$ giving a typical peak-density $\rho_0 \simeq 10^{11} \, \mathrm{cm}^{-3}$. The temperature is of the order of $200\,\mu\mathrm{K}$, measured by a
standard time-of-flight (TOF) technique. Our sample is also characterized
by its on-resonance optical thickness $b_0$. This quantity is inferred from the measurement of the transmission of a small probe beam going through the center of the atomic cloud as a function of the frequency detuning on the $|3\rangle \to |4'\rangle$ D$_2$ hyperfine transition as shown on Fig.\,\ref{fig:setup}b (frequency $\omega_0$, wavelength $\lambda = 780.24\,\mathrm{nm}$, linewidth $\Gamma/2\pi = 6.07\,\mathrm{MHz}$) \cite{Kashanian2016a}.

To measure the intensity correlations of the scattered light, we use the following time sequence. First, the MOT is loaded from a background vapor for 30 ms, followed by a compression stage of 35 ms. Then, the MOT trapping beams and the magnetic field gradient are switched off allowing for a 2-ms free expansion. Next, we apply two pulses of a weak probe beam with a waist of 250 $\mu\mathrm{m}$, linearly polarized and detuned by $\delta = \omega_\mathrm{L} - \omega_0$ with $\delta>0$, $\omega_\mathrm{L}$ being the
laser frequency. This probe beam is delivered by a distributed-feedback (DFB) laser, amplified by a tapered amplifier. Although it is known that DFB lasers have strong frequency noise~\cite{Kraft:2005,Kashanian2016a}, it does not affect our measurements of intensity correlations~\cite{Laser_noise_g2}.
The duration of each probe pulse is 100 $\mu\mathrm{s}$,
corresponding to a time-window where the intensity correlations are recorded, separated by a 50-$\mu\mathrm{s}$
pulse of repumper and 50 $\mu\mathrm{s}$ of free expansion. From one
cycle to the other, atoms are recycled and we reach a steady state with $N \simeq 3
\times 10^{9}$ atoms and $b_0 = 105 \pm 4$, measured between each acquisition of the $g^{(2)}$ function.

The setup for measuring $g^{(2)}(\tau)$ is represented in Fig.~\ref{fig:setup}. The scattered light is
collected by a single mode fiber (SMF) at a distance $L = 40\,\mathrm{cm}$ with an angle of $\theta = 41^{\circ}$
from propagation direction of the probe beam. Only one polarization is coupled into the SMF thanks to a polarizing
beam splitter (PBS). The combination of the SMF and the PBS ensures the condition of maximum spatial coherence, $\beta =
1$. The coupled light is then split using a 50:50 fiber beam splitter and finally detected on two
single-photon avalanche photodiodes (APDs, SPCM-AQRH from Excelitas Technologies), gated on the probe pulses.
Time tags,
with a resolution of 162 ps, are finally obtained from a multichannel time-to-digital converter (TDC, ID800 from
IDQuantique) and sent to a computer to record the histogram of the coincidences. Number of counts during the 200\,$\mu$s total probe pulse range from 7 to
$40$, well above the dark count rate ($< 0.02$ during the total probe pulse). Typical integration times
range from a few hours to a few days.

To reconstruct the $g^{(2)}$ function, we need to normalize the recorded coincidences. For delays much larger than the coherence time, $g^{(2)}(\tau)$ is expected to be flat and equal to 1. However, in this range, because of the finite time-window of the measurement, the number of recorded coincidences decreases linearly with the delay and reaches 0 at the pulse
duration. The $g^{(2)}$ function is thus obtained by dividing the data by the fitted slope at large delays. The detection scheme and the normalisation
were tested by measuring the light scattered by a piece of paper at rest illuminated by the probe laser, giving $g^{(2)}(\tau) = 1$ within the statistical uncertainty.

Exploiting the narrow atomic resonance, we can easily tune the optical thickness of the sample
\begin{equation}
b(\delta) = \frac{b_0}{1 + 4\delta^2/\Gamma^2},
\end{equation}
by varying the detuning $\delta$ of the probe. We can thus explore the transition from the single-scattering regime, where $b(\delta) \ll 1$, to the multiple-scattering regime, $b(\delta) \gg 1$, in principle up to $b \sim 100$.
However, one limitation is the amount of light scattered in the detection direction, which decreases with $b$ for $b>1$~\cite{Labeyrie2004}. To keep a reasonable integration time, we have thus limited our investigation to $b(\delta) \leq 2$. Note that the intensity of the probe is adapted to each detuning in order to keep the saturation
parameter constant at $s \simeq 5 \times 10^{-2}$, with $s=(I/I_\mathrm{sat})/(1+4\delta^2/\Gamma^2)$ and $I_\mathrm{sat} \simeq 3.7$~mW/cm$^{2}$. This low value of $s$ ensures that inelastic scattering is negligible.


\section{Results and Discussion} \label{Sec:Results}

\begin{figure*}[t]\centering
	\includegraphics{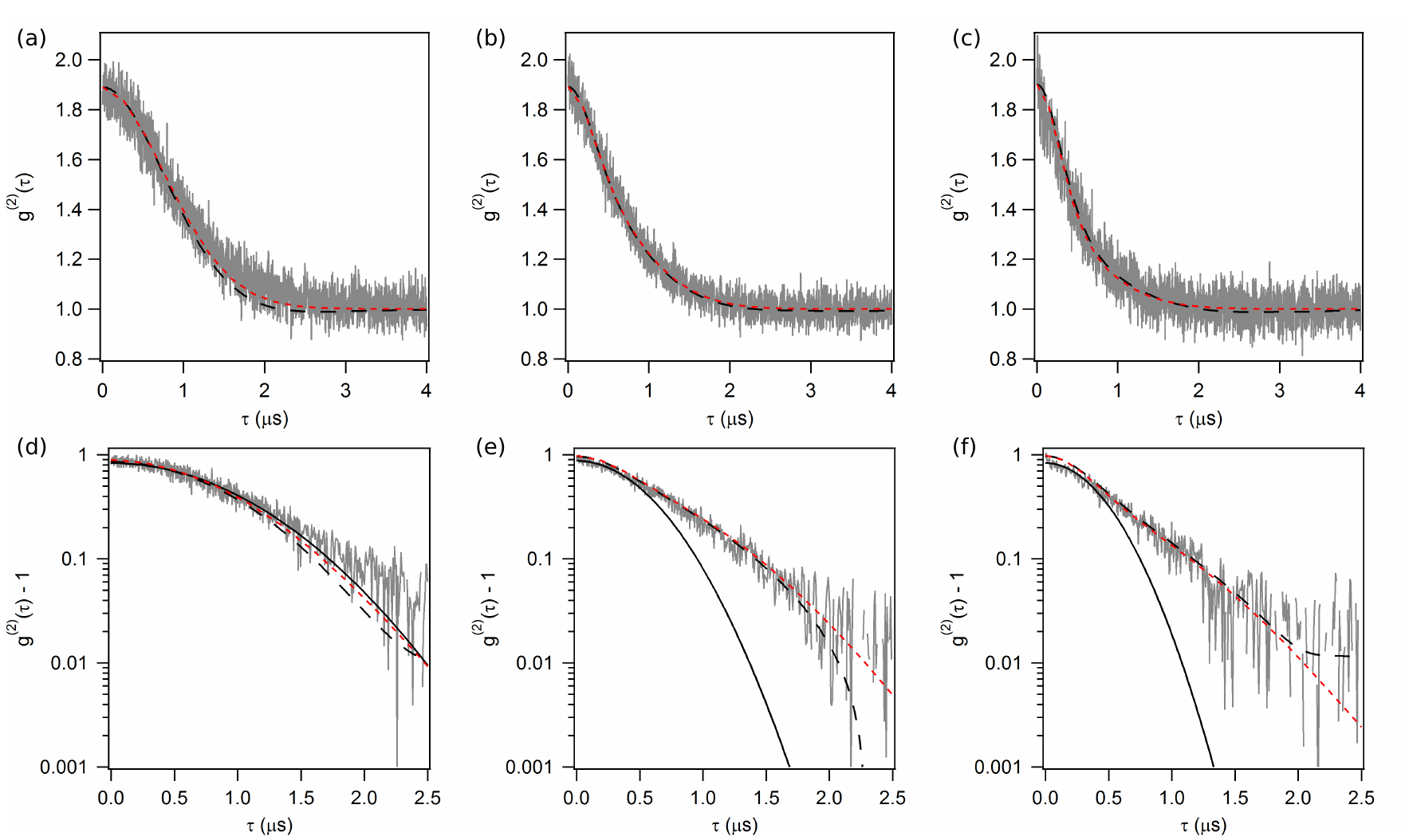}
	\caption{Intensity correlation function $g^{(2)}(\tau)$ with $b_0 \simeq 105$ and different detunings $\delta/\Gamma = \{20, 5, 3\}$ from left to right, corresponding to different optical thicknesses, (a)-(d)
	$b(\delta) = 0.07$, (b)-(e) $b(\delta) = 1$ and (c)-(f) $b(\delta) = 2$. Grey continuous curve: experimental data; red dotted dashed line: random walk simulations; black dashed line: coupled-dipole simulations.
	The contrast $g^{(2)}(0) - 1$ obtained in the simulations is equal to one. But to take into account the loss of contrast due to Raman scattering, the curves are renormalized by the mean value of the experimental $g^{(2)}$ function from $-30$\,ns to $+30$\,ns. Top: curves in linear scale. Bottom: zoom on the first $2.5\,\mu\mathrm{s}$ in semi-log scale. A quadratic behavior is added (black solid line) to guide the eyes. The difference between the model and the experimental results at large $\tau$ in (e) and (f) is due to numerical uncertainties.}
	\label{fig:g2_with_b}
\end{figure*}

We present in Fig.~\ref{fig:g2_with_b} several measurements of $g^{(2)}(\tau)$ for different $b(\delta)$,
from the single-scattering to the multiple-scattering regime. From these experimental curves, three observations can be made, as detailed in the following.

First, a systematic analysis of the contrast, defined as $\mathcal{C} = g^{(2)}(0)-1$, with respect to the detuning shows that it is slightly lower than the ideal value of one. We have checked that this loss of contrast could not be explained by any correlation artefacts (due to the light scattered by the
vacuum cell or by the hot vapor for example) or by the dark count of the detectors. We attribute this decrease to spontaneous
Raman scattering between the two hyperfine ground states, $|3\rangle \to |2\rangle$, via off-resonant excitation of the
$|3'\rangle$ and $|2'\rangle$ levels (see Fig.~\ref{fig:setup}b). Since the two ground states are separated by $3\,\mathrm{GHz}$, the interference between Rayleigh- and Raman-scattered
light induces a beating that is too fast to be resolved with the $\sim 350$~ps time resolution of our APDs. This leads to a time
averaging and thus a reduction of the contrast, as already seen in~\cite{Dussaux2016} with hot vapors. As the detuning from the $|4'\rangle$ level increases, the
Rayleigh scattering decreases, whereas the Raman scattering is almost constant. Thus, the relative weight of Raman scattering
increases and the expected contrast slightly decreases. Our signal to noise ratio is not good enough to observe this behavior, but the experimental loss of contrast roughly corresponds to the expected one, with a mean value of the order of 10$\%$ for our experimental parameters. Finally, Raman scattering between Zeeman states does not contribute to reducing the contrast. Indeed, in the worst case of a non-perfect degeneracy, it would result in a very slow beating which would be resolved by our detection scheme.

Second, increasing the optical thickness leads to a narrowing of the intensity correlation function, i.e., to a decrease of the temporal coherence. This is a direct signature of the frequency redistribution induced by multiple-scattering. Each scattering event is indeed associated to a frequency shift $\delta\omega = \vec{\Delta{}k}\cdot\vec{v}$ due to Doppler effect, where $\vec{\Delta k}$ is the
difference between the wavevectors of the scattered photon and of the incident one, and $\vec{v}$ is the
velocity of the scattering atom. Since all velocity directions are equiprobable, $\delta\omega$ is a centered random variable~\cite{footnote1}. As a consequence this frequency redistribution induces a diffusion in frequency space, leading in average to a broadening of the spectrum, and correspondingly a decrease of the temporal correlation with the number of scattering events, and thus with the optical thickness. Note that previous signatures of this effect have been indirectly observed in radiation-trapping experiments~\cite{Labeyrie2003,Labeyrie2005,Pierrat:2009}. This reduction of the coherence time with multiple-scattering is the key idea of DWS, since it corresponds to an increased sensitivity to the motion of scatterers. It should also be noted that contrary to previous reports~\cite{Article:Bali1996,Stites2004}, multiple-scattering does not induce here any reduction of the contrast of $g^{(2)}(\tau)$, because we only have here coherent (Rayleigh) scattering.

\begin{figure}\centering
	\includegraphics[width=86mm]{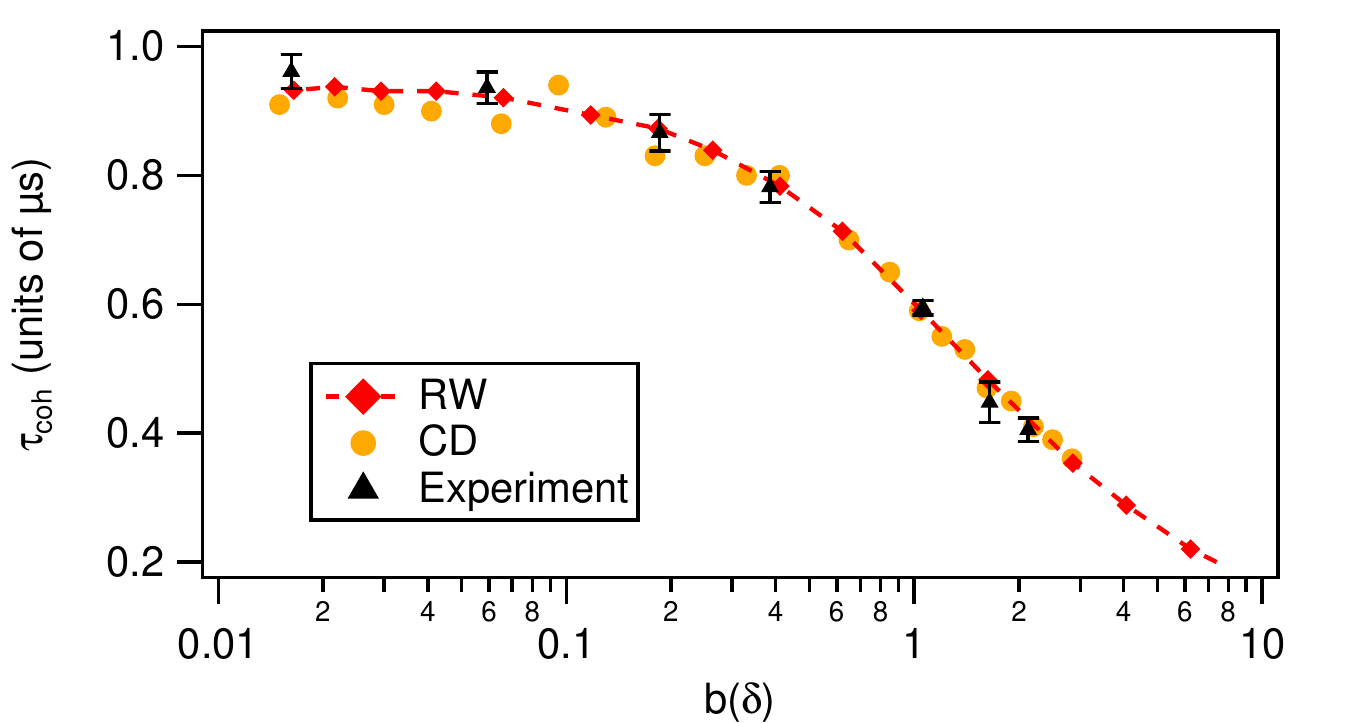}
	\caption{Coherence time, defined as the HWHM of $g^{(2)}(\tau)$, as a function of the detuning-dependent optical thickness. The black triangles with error bars are extracted from experimental data, the red diamonds from the random walk simulations and the orange circles from the coupled-dipole simulations.}
      \label{fig:CompExpSimu}
\end{figure}

A more detailed study of the evolution of the coherence time $\tau_{\text{coh}}$, defined here as the half width at half maximum (HWHM) of the $g^{(2)}$ function, as a function of the optical thickness $b(\delta)$ is reported in Fig.~\ref{fig:CompExpSimu}. The HWHM extracted from the experimental $g^{(2)}$ function correspond to the black triangles. It shows a plateau at very small $b$ and a decrease as soon as $b$ is not much smaller than unity. The plateau is due to Doppler broadening in the single-scattering regime. It is directly related to the temperature of the atomic sample but also strongly depends on the direction of observation $\theta$ from the incident wave vector. The Doppler shift for each scattering event is indeed given by $\delta\omega(\theta) = \vec{\Delta{}k}\cdot\vec{v} = k[v_\parallel (\cos \theta -1) + v_\perp \sin \theta]$, where $v_\parallel, v_\perp$ are the longitudinal and transverse components of the atomic velocity along the direction defined by the incident wave vector, and $k = 2\pi/\lambda$. For cold atoms, $\langle v_\parallel^2 \rangle = \langle v_\perp^2 \rangle = k_{\text{B}}T/M$, $M$ being the atomic mass of $^{85}\mathrm{Rb}$ and $k_{\text{B}}$ the Boltzmann
constant, which gives a Doppler width $\Delta\omega(\theta) = k\sqrt{2(1-\cos\theta)k_{\text{B}}T/M}$~\cite{footnote1}. Since the atomic velocity is given by the Maxwell-Boltzmann distribution, the Doppler-broadened spectrum is Gaussian:
\begin{equation}
S(\omega_\mathrm{L},\theta) \propto e^{-(\omega_\mathrm{L}-\omega_0)^2/2\Delta\omega^2(\theta)}. \label{eq:S1}
\end{equation}
Therefore, in the single-scattering limit, the first order coherence function $g^{(1)}(\tau)$, which is the Fourier transform of the optical spectrum, is also Gaussian:
\begin{equation}
g^{(1)}(\tau,\theta) = e^{-\tau^2 (1-\cos \theta) / 4 \tau^2_{\mathrm{c}}}, \label{eq:g1_1}
\end{equation}
with $\tau_{\mathrm{c}}^{-1} = 2k \sqrt{k_\mathrm{B}T/M}$ and, according to the Siegert relation~\eqref{eq:Siegert}, so is $g^{(2)}(\tau)$:
\begin{equation}
g^{(2)}(\tau) = 1 + e^{-\tau^2 (1-\cos \theta) / 2 \tau^2_{\mathrm{c}}}. \label{eq:g2_1}
\end{equation}
Taking the data at the lowest optical thickness $b = 0.02$, we extract a HWHM $\tau_{\text{coh}} = (961 \pm 30)\,\mathrm{ns}$, corresponding to a temperature of $(245 \pm 15)\,\mu\mathrm{K}$.

Finally, the third observation featured on Fig.~\ref{fig:g2_with_b} is the progressive change of shape of the $g^{(2)}$ function. At low optical thicknesses, $g^{(2)}$ is well fitted by a Gaussian. As expected and explained before, this is obtained due to the Maxwell-Boltzmann distribution of the atomic velocity but also because we can assume that the atoms undergo purely ballistic motions during the whole scattering process. It is important to note that, for standard DWS performed on suspended particles in a liquid, this regime is hardly reached. The interactions with the fluid induce in particular a long-time diffusive motion for the particles and thus an exponential decay for $g^{(2)}$\,\cite{Maret1987, Pine1988, Weitz1989}.

At short time scale, $g^{(2)}$ remains Gaussian with a width that decreases as $b$ is increased, as shown on the insets of Fig.\,\ref{fig:g2_with_b}. On the other hand, in the multiple-scattering regime and for larger time-scale, the shape clearly deviates from a Gaussian, with a decrease that gets closer to linear in semi-log scale instead of quadratic, and becomes more complex to infer analytically. In order to understand this change, we have performed numerical simulations to probe the role of multiple-scattering, based on the two following models.

\begin{figure}\centering
	\includegraphics[width = 86mm]{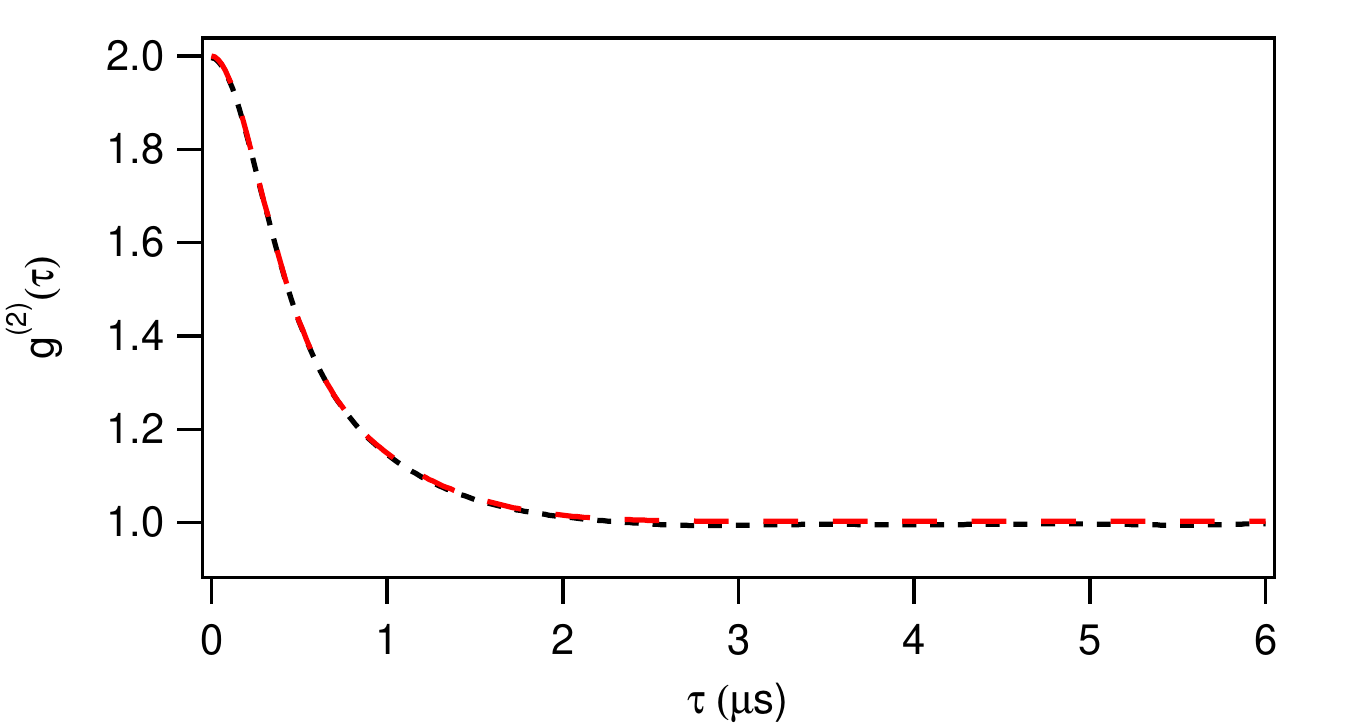}
	\caption{Verification of the Siegert relation with a cloud of $N=5000$ atoms, $b_0=2.8$, $b(\delta) = 2$ and $\theta = 41^\circ$. Black dashed line: $g^{(2)}(\tau)$; red dashed line: $1 + |g^{(1)}(\tau)|^2$.
The Siegert relation is well satisfied.}
      \label{fig:Siegert}
\end{figure}

The first model, known
as the coupled-dipole model (CD)~\cite{Svidzinsky:2010,Courteille:2010,Bienaime:2011,Bienaime:2013}, describes the sample as an ensemble of identical atoms driven by a laser field. Atoms are modeled as point-like dipoles interacting via the light in the scalar approximation. Their position changes in time following the velocity distribution corresponding to the temperature, and they are spatially distributed with a Gaussian shape
that mimics the geometry of the cloud. In the recent years, this model has been widely used in the context of single-photon superradiance and subradiance~\cite{Scully:2006,Scully:2009,Guerin2016,Araujo2016,Roof2016}. It is valid at low intensity, when the saturation parameter $s \ll 1$, but it includes coherent and cooperative effects. Since this coherent model allows computing the time-dependent radiated field of each dipole, it provides access to both the $g^{(1)}(\tau)$ and $g^{(2)}(\tau)$, so it can be used to checked the validity of the Siegert relation, as demonstrated in Fig.\,\ref{fig:Siegert} for $b = 2$. The use of the CD model is however limited to a few thousands atoms and thus to a maximum resonant optical thickness of $b_0 \sim 10$ for dilute samples. Still, by adapting the detuning, the same $b(\delta)$ as in the experiment can be simulated. Simulations of the CD model reveal that the intensity correlation function indeed depends only on $b(\delta)$.

The second model is a Monte-Carlo simulation modeling
an incoherent random walk (RW) of photons inside the atomic cloud. The step of a photon sent into the medium is
randomly sampled taking into account a step-length distribution due to the Gaussian density profile of the cloud and the dependent-detuning isotropic scattering cross-section. Doppler effect is included in the model by adding a frequency shift after each scattering event, randomly sampled from the velocity
distribution. It is then possible to select all photons exiting at a particular angle and to get the output spectrum of the scattered light $S(\omega_\mathrm{L},\theta)$, which corresponds to the Fourier transform of the $g^{(1)}$ function. We can finally access the $g^{(2)}$ function by assuming the Siegert relation~\eqref{eq:Siegert}, validated by the CD model.

The RW simulations are useful to understand the shape of the intensity correlation function. Indeed, these simulations give, for each photon exiting the medium, the optical frequency, the outgoing angle and the number of undergone scattering events. We then extract, at a particular angle, the number of photons escaping the cloud after $n$ scattering events from which we construct the distribution $P(n)$. Besides, from the optical frequency, one also obtains the corresponding optical spectrum $S_n(\omega_\mathrm{L},\theta)$ and its Fourier transform $g_n^{(1)}(\tau,\theta)$. The total $g^{(2)}$ function is finally given by:
\begin{equation}
g^{(2)}(\tau) = 1+ \left|\sum_{n> 0} P(n) g_n^{(1)} (\tau,\theta)\right|^2.
\end{equation}
This expression is useful to understand the role of the different scattering orders on the shape of the $g^{(2)}$ function, which depends on the individual shape of $g_n^{(1)}$ as well as $P(n)$, see Fig.~\ref{fig:g2_with_b}.

\begin{figure}\centering
	\includegraphics[width=86mm]{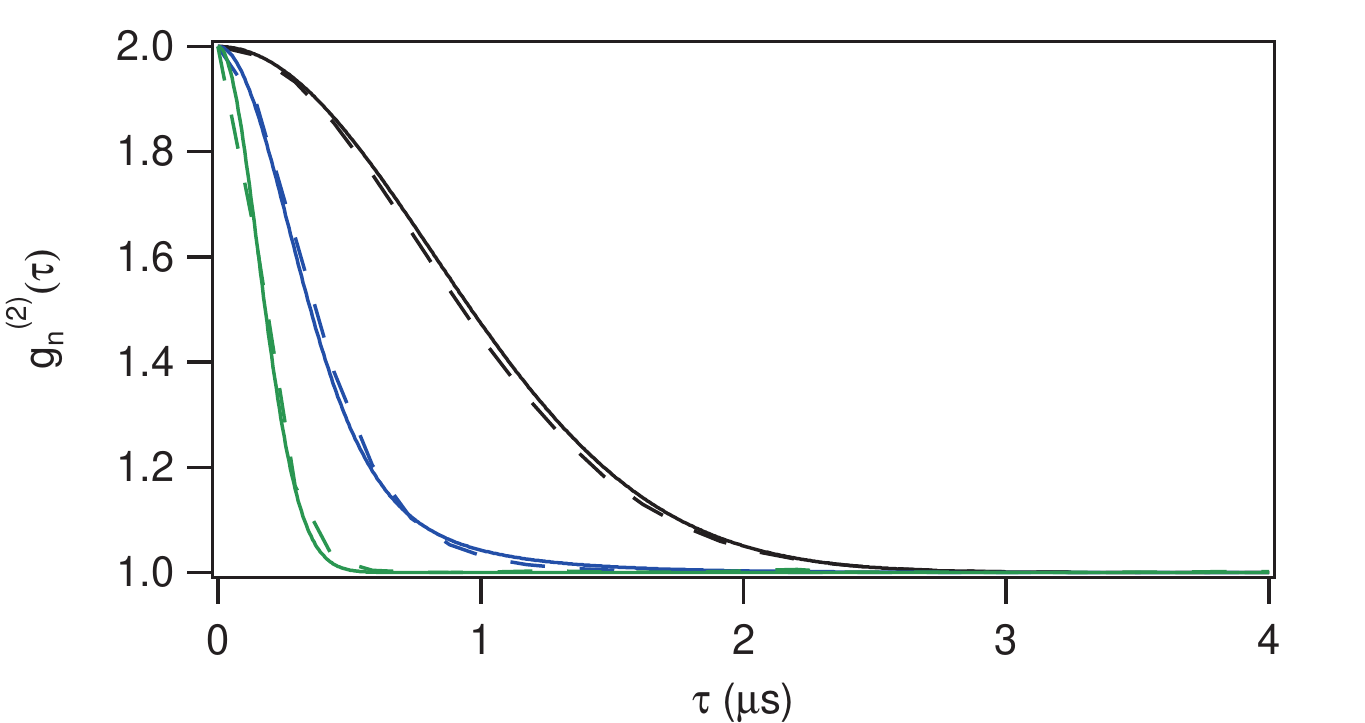}
	\caption{Intensity correlation function $g_n^{(2)}(\tau,\theta = 41^\circ)$ for photons that have been scattered $n$ times before escaping. Solid line: obtained using the analytical expressions \eqref{eq:g1_1}, \eqref{eq:g2_1} and from \eqref{eq:g2_ana} to ~\eqref{eq:gn_ana}. Dashed line: extracted from the RW simulations with $b(\delta) = 2$. In diminishing order of width: $n = 1$ (black), $n = 2$ (blue) and $n = 7$ (green).}
      \label{fig:g2_analytique}
\end{figure}

While an analytical expression of $P(n)$ is hardly accessible, we have checked that, for small $b$ ($b\lesssim 2$), $g_n^{(1)}(\tau,\theta)$ can be obtained analytically. For one scattering event, one gets the spectrum and the first order correlation function given by Eq.\,(\ref{eq:S1}) and Eq.\,(\ref{eq:g1_1}). For two scattering events, let us first consider the case of a first photon scattered in the $\theta_1$ direction. Since the measurement is done in the $\theta$ direction, the second photon must be scattered at the angle $\theta-\theta_1$ compared to the first one. The spectrum is then given by the convolution of the two individual Doppler spectra $S(\omega_L,\theta_1) \ast S(\omega_L,\theta-\theta_1)$. Taking the Fourier transform, the corresponding $g^{(1)}$ function is $g_1^{(1)} (\tau,\theta_1) \times g_1^{(1)} (\tau,\theta-\theta_1)$. Finally, the total $g_2^{(1)}$ is the sum over all the possible angles:
\begin{eqnarray}
g_2^{(1)} (\tau,\theta) &=& \int_0^{2\pi} g_1^{(1)} (\tau,\theta_1) g_1^{(1)} (\tau,\theta-\theta_1) d\theta_1 \label{eq:g2_2}\\
&=& e^{-\frac{2\tau^2}{4 \tau^2_{\mathrm{c}}}} I_0\left(\frac{\tau^2}{4\tau^2_{\mathrm{c}}} \sqrt{2(1+\cos \theta)} \right), \label{eq:g2_ana}
\end{eqnarray}
with $I_0$ the modified Bessel function of the first kind. While in the single-scattering regime the $g_2^{(1)}$ is a Gaussian function, its shape is more complex in the double scattering regime. The principle of the calculation is the same when the number of scattering events is increased and one gets:
\begin{eqnarray}
g_n^{(1)} (\tau,\theta) &=& \int_0^{2\pi} \left[\prod_{i = 1}^{n-1}g_1^{(1)} (\tau,\theta_i)\right] g_1^{(1)} \left(\tau,\theta-\sum_{j = 1}^{n-1}\theta_j\right) \nonumber\\
&& d\theta_1...d\theta_{n-1},
\end{eqnarray}
When $n$ gets larger, the contribution of $\theta$ is blurred. Moreover the Gaussian part decreases more rapidly than the Bessel function and one recovers the Gaussian shape:
\begin{equation}
g_n^{(1)} (\tau,\theta) \underset{n\to \infty}{\longrightarrow} e^{-\frac{n\tau^2}{4 \tau^2_{\mathrm{c}}}} \label{eq:gn_ana},
\end{equation}
with a HWHM that depends on the number of scattering events as $1/\sqrt{n}$. As said before, the $g_n^{(2)}$ functions are calculated using the Siegert relation. The analytical expressions are compared to the ones obtained by the RW simulations in Fig.\,\ref{fig:g2_analytique}, showing a good agreement between them.

To compare the experimental results and the simulations, we first extract the coherence time. The results are plotted in Fig.~\ref{fig:CompExpSimu} with the two models. The RW simulations are performed with the actual parameters of the experiment and the CD simulations use a lower $b_0$ and correspondingly lower detunings to match the same $b(\delta)$. The only free parameter is the temperature, which is chosen in the simulations in order to match the one extracted from the $\tau_\mathrm{coh}$ measured with the lowest $b$. Despite the differences between the two models, there is a remarkable agreement between them and with the experimental data.

Finally, the shape of $g^{(2)}(\tau)$ obtained experimentally is compared with the simulations in Fig.~\ref{fig:g2_with_b}.
Both models are based on a two-level atom, the contrast is thus maximal and equal to one. However, in order to
compare on the same figure the numerical and the experimental curves, the loss of contrast due to Raman scattering has to be taken into account. The numerical ones are thus renormalized by the experimental contrast, evaluated by taking the mean value of the experimental $g^{(2)}$ function from $-30$\,ns to $+30$\,ns. We see a good agreement between all curves, showing that the effect of the multiple-scattering is well described by both the CD and the RW models.



\section{Conclusion} \label{Sec:Conclusion}

To summarize, we have precisely measured the temporal intensity correlation function $g^{(2)}(\tau)$ of light scattered by cold atoms in ballistic motion, a regime hardly reached in standard DWS measurements. Different regimes of light transport have been investigated, from the single-scattering to the multiple-scattering regime. We have observed a decrease of the width of $g^{(2)}(\tau)$ as we enter the multiple-scattering regime and we have evidenced the corresponding change of shape in $g^{(2)}(\tau)$. Numerical simulations, based on the coupled-dipole model or on random walk simulations, have been performed to further understand the role of multiple-scattering of light in intensity correlations. A remarkable agreement is obtained between the two simulations and with experimental data.

DWS is a powerful technique that has been widely used in soft condensed matter but it can find many applications in other field of research, ranging from atomic physics to astrophysics. In hot atomic vapors, DWS might allow to study time dependance in Levy flights\,\cite{Mercadier2009, Baudouin2014} and the connection between L\'evy flights and L\'evy walks. In cold atomic samples, it might be used to quantify the friction induced by Doppler or sub-Doppler cooling. Anomalous diffusion, as expected close to the decrochage could be investigated with an alternative approach to that used in\,\cite{Jersblad2004, Douglas2006}. DWS might also prove to be a powerful tool to investigate phase transitions, as for instance Anderson localization of light by cold atoms\,\cite{Skipetrov2015, Celardo2017}, and to study the impact of Dicke type super- and subradiance\,\cite{Guerin2016,Araujo2016,Roof2016} on fluctuations of scattered or emitted light. Finally, intensity correlations are also a well known tool to distinguish classical from quantum optics\,\cite{Glauber:1963}. This will allow us to probe the onset of gain and random lasing in our cloud of cold atoms\,\cite{BaudouinNatPhys2013}. Gain and lasing being also expected in astrophysical systems\,\cite{Letokhov2009}, the recent success of intensity correlations measured on three bright stars \cite{HBT:nous} opens this quantum eye to lasing in astrophysics.

\section*{Aknowledgements}

We thank Antoine Dussaux for the TDC programs and S. Vartabi Kashanian for the early stage of the DWS experiment. A.E. is supported by a grant from the DGA. R. B. benefited from Grants from São Paulo Research Foundation
(FAPESP) (Grant Nos. 2015/50422-4  and 2014/01491-0). R. B., M. F and R. K. received support from project CAPES-COFECUB (Ph879-17/CAPES 88887.130197/2017-01). The Titan X Pascal used for this research was donated by the NVIDIA Corporation. We wish to thank the F\'ed\'eration Doeblin for financial support.

%


\begin{thebibliography}{56}
\bibitem [{Las()}]{Laser_noise_g2}%
  \BibitemOpen
  \href@noop {} {}\bibinfo {note} {The laser frequency noise does not affect
  the $g^{(2)}$ function as soon as the typical path length of the photon
  inside the atomic sample is much lower than the laser coherence
  length.}\BibitemShut {Stop}%
\end{thebibliography}

\end{document}